# Domain structure in CoFeB thin films with perpendicular magnetic anisotropy

Michihiko Yamanouchi[1], Albrecht Jander[2,3], Pallavi Dhagat[2,3], Shoji Ikeda[1,3], Fumihiro Matsukura[1,3], and Hideo Ohno[1,3]

[1] *Center for Spintronics Integrated Systems, Tohoku University, Katahira 2-1-1, Aoba-ku, Sendai 980-8577, Japan*

[2] *School of Electrical Engineering and Computer Science, Oregon State University, Corvallis, Oregon 97331-3211, USA*

[3] *Laboratory for Nanoelectronics and Spintronics, Research Institute of Electrical Communication, Tohoku University, Katahira 2-1-1, Aoba-ku, Sendai 980-8577, Japan*

**Abstract**

Domain structures in CoFeB-MgO thin films with a perpendicular easy magnetization axis were observed by magneto-optic Kerr-effect microscopy at various temperatures. The domain wall surface energy was obtained by analyzing the spatial period of the stripe domains and fitting established domain models to the period. In combination with SQUID measurements of magnetization and anisotropy energy, this leads to an estimate of the exchange stiffness and domain wall width in these films. These parameters are essential for determining whether domain walls will form in patterned structures and devices made of such materials.

**I. Introduction**

The interface between CoFeB and MgO in CoFeB films generates a large magnetic anisotropy in the direction perpendicular to the film plane [Endo 2010, Ikeda 2010]. Although the presence of perpendicular interface anisotropy has been previously reported in CoFeB-MgO [Hosomi 2007, Nistor 2009, Yakata 2009, Shimabukuro 2010], realization of magnetic tunnel junctions (MTJs) with a perpendicular easy axis in the CoFeB-MgO system is of particular interest since these materials are now commonly used for MTJs with high magnetoresistance [Ikeda 2010]. The CoFeB-MgO system with a perpendicular easy axis is of advantage to maintain thermal stability in nano-scale devices and has also been shown to result in lower threshold currents for spin transfer torque switching in MTJs [Ikeda 2010]. Moreover, unlike other recently reported multilayers with perpendicular anisotropy [Jung 2010], an additional metallic layer adjacent to the CoFeB is not needed to obtain a perpendicular easy axis. Such metallic adjacent layers reduce spin torque effects in devices utilizing domain wall motion where current flows in the film plane because they shunt the current. Indeed, low threshold current to induce domain wall motion has recently been reported in CoFeB/MgO nanowires with perpendicular anisotropy [Fukami 2011]. Therefore the CoFeB-MgO system with perpendicular anisotropy is very attractive for application to nonvolatile high-speed domain wall devices utilizing current-induced domain wall motion, in addition to MTJs for magnetoresistive random access memories [Parkin 2004, Numata 2007, Kawahara 2008, Matsunaga 2008].

Here we investigate the properties of CoFeB thin films with a perpendicular magnetic anisotropy by observing temperature dependence of magnetic domain structure [Saratz 2010a, Saratz 2010b] and comparing the domain structure in demagnetized states with that predicted by established domain models [Málek 1958, Kooy 1960, Kaplan 1993]. The spatial period of the stripe domains that form in the films provides a measure of the domain wall energy, $\gamma_w$. This, in combination with superconducting quantum interference device (SQUID) magnetometry of the magnetization and magnetic anisotropy yields an estimate of the exchange stiffness, $A_s$, and domain wall width, $\delta_w$, in the films. These parameters determine conditions for domain wall stability in devices utilizing domain wall motion or the formation of unwanted domains in MTJs [Kitakami 2009].

**II. Experimental Procedure**

All samples were deposited on thermally oxidized Si (001) substrates using RF magnetron sputtering at room temperature. The stack consists of substrate/Ta(5)/Ru(10)/Ta(5)/Co$_{20}$Fe$_{60}$B$_{20}$($t_{CoFeB}$ = 0.9-1.3)/MgO(1)/Ta(2) (numbers are nominal film thickness in nanometers determined from the deposition rate). After deposition of the multilayers, each wafer was cut into two pieces and one was annealed in a vacuum for 1 hour under a perpendicular magnetic field of 0.4 T at 350°C. Both the as-deposited and annealed samples were cleaved into 4 mm squares for SQUID measurements and smaller pieces, a few mm$^2$, for domain observations.

**III. Results and Discussion**

Magnetization versus external magnetic field (*M−H*) curves of sample A (as-deposited sample with the thickness of CoFeB, $t_{CoFeB}$ = 1.1 nm) and sample B (annealed sample with $t_{CoFeB}$ = 1.3 nm) were measured by SQUID at various temperatures (10−300 K). Figure 1(a) shows the *M−H* curves for sample A at 300 K under in-plane and out-of-plane magnetic fields. The nearly linear, non-hysteretic response to in-plane fields is characteristic of perpendicular magnetic anisotropy. From the out-of-plane *M-H* curves, the value of coercive field is less than 1 mT. Similar *M-H* curves are obtained with sample B at 300 K and the value of coercive field is also less than 1 mT. The saturation magnetization $M_s$ and perpendicular magnetic anisotropy energy density $K = K_{eff} + \mu_0 M_s^2/2$ are extracted from the *M−H* curves, where $K_{eff}$ is an effective perpendicular anisotropy energy density. The value of $M_s$ is determined by the average of *M* over a range of magnetic fields, 1 T ≤ $|\mu_0 H| \leq 2$ T. The value of $K_{eff}$ is obtained from the area of triangular region between $M_s$ and the in-plane $M-H$ curves. Thus the value of $K_{eff}$ includes higher order contributions of uniaxial anisotropy. Figures 1(b) and 1(c) show the temperature, $T$, dependence of $M_s$ and $K$ for samples A and B, respectively. Assuming that the bulk crystalline anisotropy is negligible, the interface magnetic anisotropy energy densities, $K_i$, for samples A and B at 300 K are 0.49 mJ/m$^2$ and 1.4 mJ/m$^2$, respectively. These values are comparable to those calculated from the values of $M_s$, $K_{eff}$, and $t_{CoFeB}$ for similar samples obtained in previous work [Ikeda 2010, Yamanouchi 2011]. For both of the samples, $M_s$ and $K$ increase at lower temperatures. Since domain structures are determined by the trade-off between demagnetization energy and domain wall surface energy, which depend on the magnitude of $M_s$ and $K$, temperature dependent structural changes of domains are anticipated for both of the samples. The magnitudes of $M_s$ and $K$ of sample B are larger than those of sample A, likely due to the crystallization of CoFeB triggered by a decrease of boron concentration during annealing [Hayakawa 2005].

For magnetic domain observation, the samples were placed in an optical cryostat and the domain structure was observed by polar magneto-optic Kerr-effect (MOKE) microscopy using 546 nm illumination and 20× magnification. The maximum size of the MOKE image is 192×256 $\mu$m$^2$, limited by the capacity of the CCD camera and the magnification of the microscope. Before capturing the domain image, at each temperature, the samples were ac-demagnetized using an alternating perpendicular magnetic field with exponentially decaying amplitude starting from 20 mT. To enhance the image contrast, differential images between ac-demagnetized and remanent states after applying a large enough field to saturate magnetization were taken. Representative selections of domain images for samples A and B at various temperatures (10, 200, and 300 K) are shown in Fig. 2(a)-(c) and Fig. 2(d)-(f) respectively. The bright and dark regions correspond to out-of-plane magnetization pointing up and down respectively. At $T \geq 100$ K, the domains were quite mobile, quickly settling into the labyrinthine patterns. However, at $T \leq 50$ K, the domain walls were strongly pinned and the domain structure may not represent the lowest energy configuration. Rather, at low temperatures, the pattern reflects the distribution of domain nucleation sites because magnetization reversal of the observed region was dominated by nucleation of many domains during the ac-demagnetization procedure.

It should be noted that domains as large as a few hundred μm, which is comparable to the maximum size of the MOKE images, were observed for the as-deposited samples with $t_{CoFeB}$ = 0.9 nm and 1.0 nm and the annealed samples with $t_{CoFeB}$ = 0.9-1.2 nm at room temperature. This is due to the higher domain wall energy resulting from the greater influence of the interface anisotropy in the thinner magnetic films and is consistent with the domain theory discussed below. However, the limited field of view of the MOKE microscope prohibits a quantitative analysis of these large domain patterns. For thicker, as-deposited samples with $t_{CoFeB}$ = 1.2 and 1.3 nm, no perpendicular domain structures are detected at room temperature. Based on the measured values of $M_s$ and $K_i$ for samples A and B, the in-plane shape anisotropy is expected to dominate the perpendicular interface anisotropy for films thicker than $t_{CoFeB}$ = 1.23 and 1.39 nm in as-deposited and annealed samples, respectively. Thus, only a limited range of thicknesses is amenable to domain analysis, in thinner films the domains are too large to measure and in thicker films no domains are detected due to lack of perpendicular anisotropy. Moreover, domain structures in samples A and B cannot be compared since normalized CoFeB thicknesses of samples A and B with the critical thicknesses are different (0.89 and 0.94 for samples A and B).

To quantify the observed domain structures, domain images are analyzed using two-dimensional fast Fourier transforms (FFT). The processed images at 10 K and 300 K for sample A are shown in Figs. 3(a) and (b). The color scale is normalized with black indicating the largest amplitude and white indicating zero amplitude of the spatial frequency components. There are no apparent periodic patterns at 10 K, which suggests that randomly distributed pinning is restricting the formation of regular domain patterns. On the other hand, periodic and nearly isotropic patterns are formed at 300 K as predicted by domain theories [Málek 1958, Kooy 1960, Kaplan 1993]. These results show that pinning of domain wall is unimportant for the formation of demagnetized state in CoFeB/MgO system at $T \geq 100$ K. Figure 4(a) shows the amplitude of direction-independent Fourier components obtained by averaging the FFT over circles of radius 1/wavelength. The peak value is taken as the characteristic period $D_p$ of the domain pattern. In the demagnetized state, the up and down domains (bright and dark regions) are of equal size. The values of $D_p$ determined from the domain images at each temperatures ($T \geq 100$ K) in the manner mentioned above are shown in Figs. 4(c) and (d) for samples A and B respectively. The value of $D_p$ increases with decreasing temperature and peaks at around 150-200 K. This behavior of $D_p$ is in agreement with that predicted for stripe domain period in ultrathin magnetic films [Gehring 1993, Polyakova 2007]. On the other hand, $D_p$ starts to decrease at temperatures below 150 K, which may be attributed to the appearance of domain wall pinning effects on forming the lowest energy domain configuration.

For the sake of analysis, the observed labyrinthine domain structure is treated as an ideal, periodic stripe domain pattern with cross section as indicated in Fig. 4(b). For domains much larger than the thickness of the film (i.e. $D_p \gg t$, which, in the nm-thick films considered here, applies for any domain structure that can be observed optically), Kaplan and Gehring derive an analytical expression relating stripe domain period to the film thickness, $t$: $D_p = 1.91\, t\, \exp(\pi D_0/t)$ [Kaplan 1993], where $D_0 = \gamma_w/\mu_0 M_s^2$ is a characteristic length of the material determined by the domain wall surface energy $\gamma_w$ and domain magnetization, $M_s$.

Using $M_s$ measured by SQUID magnetometry and $t$ determined from the deposition rate, $\gamma_w$ is determined from the domain period by the Kaplan-Gehring equations. The temperature dependence of $\gamma_w$ for samples A and B are shown in Fig. 4(e). For both samples, $\gamma_w$ decreases almost linearly with increasing $T$, reaching 2.5 mJ/m$^2$ in sample A and 5.9 mJ/m$^2$ in sample B at 300 K. Since $\gamma_w$ must be zero at the Curie temperature, $T_c$, linear extrapolation of the values of $\gamma_w$ at $T \geq 150$ K (to avoid domain wall pinning effects) yields estimates of 728 K and 1026 K for the Curie temperatures of samples A and B respectively. The value of $T_c$ for sample B is comparable to that for Fe ($T_c \sim 1000$ K) [Chikazumi 1964]. The smaller value of $T_c$ for sample A may be related to remaining nonmagnetic boron in CoFeB.

The exchange stiffness, $A_s$, and domain wall width, $\delta_w$, may be estimated from these measurements using the following relationships, $\gamma_w = 4(A_s K_{eff})^{1/2}$ and $\delta_w = \pi\gamma_w/(4K_{eff})$ for domain walls in zero thickness limit, uniaxial materials [Schlömann 1973]. By using the measured values of $K_{eff}$ and $\gamma_w$, the values of $A_s$ for samples A and B at 300 K are found to be 8.4 pJ/m and 31 pJ/m, respectively. The values of $\delta_w$ at 300 K for samples A and B are estimated as 43 nm and 67 nm, respectively. It should be noted that the expressions for $\gamma_w$ and $\delta_w$ above are lower and upper bounds, respectively, because $\gamma_w$ and $1/\delta_w$ increase with increasing film thickness as long as the thickness remains small compared to $(A_s/K)^{1/2}$ [Schlömann 1973]. A suitable theory for the energy of domain walls in thin films dominated by interface anisotropy has not yet been developed.

The values of $A_s$ in the CoFeB samples are in the same orders of magnitude as that of Fe ($A_s \sim 20$ pJ/m) [Chikazumi 1964]. The value of $A_s$ of sample B is larger than that of sample A, which could be attributed to a smaller exchange integral, which is also suggested by the lower estimated value of $T_c$, and the smaller magnitude of $M_s$ for sample A. Although not quantifiable, it is worth mentioning that the strong dependence of $D_p$ on $t_{CoFeB}$ as well as the very large domains in the samples with thinner CoFeB are consistent with the analysis. Assuming that the values of $M_s$ and $K_i$ are the same as those of samples for the domain observation, and negligible bulk anisotropy, the predicted periods of domains, $D_p$, in the as-deposited films with $t_{CoFeB}$ = 0.9-1.0 nm and annealed films $t_{CoFeB}$ = 0.9-1.2 nm are greater than 100 μm at 300 K.

## IV. Conclusions

In conclusion, domain structures in CoFeB-MgO thin films with a perpendicular easy axis were observed by MOKE microscopy at various temperatures. At $T \geq 100$ K, domain walls were quite mobile and domains were quickly settling into labyrinthine patterns in demagnetized state. However, at $T \leq 50$ K, the domain walls were strongly pinned and the domain structure could not be interpreted by standard theories. Temperature dependence of domain wall surface energy was obtained by analyzing the spatial period of the stripe domains and fitting established domain models to the period. In combination with SQUID measurements of magnetization and anisotropy energy, this leads to an estimate of the exchange stiffness and domain width in these films. Thus determined domain wall width is comparable to the feature size of present day nanoelectronics devices. For further miniaturization of domain wall devices made of the CoFeB/MgO system, reduction of $\delta_w$ is important by decreasing demagnetization energy or increasing perpendicular anisotropy.

**Acknowledgments**

We thank I. Morita, T. Hirata, and H. Iwanuma for their technical support. This work was supported by the FIRST program from JSPS.

**References**

Endo M, Kanai S, Ikeda S, Matsukura F, Ohno H (2010), "Electric-field effects on thickness dependent magnetic anisotropy of sputtered MgO/Co$_{40}$Fe$_{40}$B$_{20}$/Ta structures," *Appl. Phys. Lett.*, vol. 96, 212503, doi: 10.1063/1.3429592.

Ikeda S, Miura K, Yamamoto H, Mizunuma K, Gan H D, Endo M, Kanai S, Hayakawa J, Matsukura F, Ohno H (2010), "A perpendicular-anisotropy CoFeB–MgO magnetic tunnel junction," *Nature Mater.*, vol. 9, pp. 721-724, doi: 10.1038/nmat2804.

Hosomi M, Yamamoto T, Higo Y, Yamane K, Oishi Y, Kano H (2007), "Progress in spin transfer torque MRAM (SpRAM) development," *Magnetics Japan*, vol. 2, pp. 606-614 [in Japanese].

Nistor L E, Rodmacq B, Auffret S, Dieny B (2009), "Pt/Co/oxide electrodes for perpendicular magnetic tunnel junctions," *Appl. Phys. Lett.*, vol. 94, 012512, doi: 10.1063/1.3064162.

Yakata S, Kubota H, Suzuki Y, Yakushiji K, Fukushima A, Yuasa S, Ando K (2009), "Influence of perpendicular magnetic anisotropy on spin-transfer switching current in CoFeB/MgO/CoFeB magnetic tunnel junctions," *J. Appl. Phys.*, vol. 105, 07D131, doi: 10.1063/1.3057974.

Shimabukuro R, Nakamura K, Akiyama T, Ito T (2010), "Electric field effects on magnetocrystalline anisotropy in ferromagnetic Fe monolayers," *Physica E*, vol. 42, pp. 1014-1017, doi: 10.1016/j.physe.2009.11.110.

Jung J H, Jeong B, Lim S H, Lee S R (2010), "Strong perpendicular magnetic anisotropy in CoFeB/Pd Multilayers," *Appl. Phys. Express*, vol. 3, 023001, doi: 10.1143/APEX.3.023001.

Fukami S, Suzuki T, Nakatani Y, Ishiwata N, Yamanouchi M, Ikeda S, Kasai N, Ohno H (2011), "Current-induced domain wall motion in perpendicularly magnetized CoFeB nanowire," *Appl. Phys. Lett.*, vol. 98, 082504, doi: 10.1063/1.3558917.

Parkin S S P (2004), *U.S. Patent*, 6834005.

Numata H, Suzuki T, Ohshima N, Fukami S, Ngahara K, Ishiwata N, Kasai N (2007), "Scalable cell technology utilizing domain wall motion for high-speed MRAM," in *2007 Symposium on VLSI Technology, Kyoto*, *Technical Digest* (IEEE, New York, 2007), pp. 232-233, doi: 10.1109/VLSIT.2007.4339705.

Kawahara T, Takemura R, Miura K, Hayakawa J, Ikeda S, Lee Y M, Sasaki R, Goto Y, Ito K, Meguro T, Matsukura F, Takahashi H, Matsuoka H, Ohno H (2008), "2 Mb SPRAM (SPin-Transfer Torque RAM) with bit-by-bit bi-directional current write and parallelizing-direction current read," *IEEE J. Solid-State Circuits*, vol. 43, pp. 109-120, doi: 10.1109/JSSC.2007.909751.

Matsunaga S, Hayakawa J, Ikeda S, Miura K, Hasegawa H, Endoh T, Ohno H, Hanyu T (2008), "Fabrication of a nonvolatile full adder based on logic-in-memory architecture using magnetic tunnel junctions," *Appl. Phys. Express*, vol. 1, 091301, doi: 10.1143/APEX.1.091301.

Saratz N, Lichtenberger A, Portmann O, Ramsperger U, Vindigni A, Pescia D (2010a), "Experimental phase diagram of perpendicularly magnetized ultrathin ferromagnetic films," *Phys. Rev. Lett.*, vol. 104, 077203, doi: 10.1103/PhysRevLett.104.077203.

Saratz N, Ramsperger U, Vindigni A, Pescia D (2010b), "Irreversibility, reversibility, and thermal equilibrium in domain patterns of Fe films with perpendicular magnetization," *Phys. Rev. B*, vol. 82, 184416, doi: 10.1103/PhysRevB.82.184416.

Málek Z and Kamberský V (1958), "On the theory of the domain structure of thin films of magnetically uni-axial materials," *Czech. J. Phys.*, vol. 8, pp. 416-421, doi: 10.1007/BF01612066.

Kooy C and Enz U (1960), "Experimental and theoretical study of the domain configuration in thin layers of $BaFe_{12}O_{19}$," *Philips Res. Rep.*, vol. 15, pp. 7-29.

Kaplan B and Gehring G A (1993), "The domain structure in ultrathin magnetic films," *J. Magn. Magn. Mater.*, vol. 128, pp. 111-116, doi: 10.1016/0304-8853(93)90863-W.

Kitakami O, Okamoto S, Kikuchi N, Shimatsu T, Mitsuzuka K, Aoi H (2009),"Bistability condition of circular nanomagnet," *Appl. Phys. Express*, vol. 2, 123002, doi: 10.1143/APEX.2.123002.

Yamanouchi M, Koizumi R, Ikeda S, Sato H, Mizunuma K, Miura K, Gan H D, Matsukura F, Ohno H (2011), "Dependence of magnetic anisotropy on MgO thickness and buffer layer in $Co_{20}Fe_{60}B_{20}$-MgO structure," *J. Appl. Phys.*, vol. 109, 07C712, doi: 10.1063/1.3554204.

Hayakawa J, Ikeda S, Matsukura F, Takahashi H, Ohno H (2005), "Dependence of giant tunnel magnetoresistance of sputtered CoFeB/MgO/CoFeB magnetic tunnel junctions on MgO barrier thickness and annealing temperature," *Jpn. J. Appl. Phys.*, vol. 44, pp. L587-L589, doi: 10.1143/JJAP.44.L587.

Gehring G A and Keskin M (1993), "The temperature dependence of the domain spacing in ultrathin magnetic films," *J. Phys. Cond. Mater.*, vol. 5, pp. L581-L585, doi: 10.1088/0953-8984/5/44/007.

Polyakova T, Zablotskii V, Maziewski A (2007), "Temperature dependence of magnetic stripe domain period in ultrathin films," *J. Magn. Magn. Mater.*, vol. 316, e139-e141, doi: 10.1016/j.jmmm.2007.02.063.

Chikazumi S (1964), *Physics of Magnetism,* New York: Wiley.

Schlömann E (1973), "Domain walls in bubble films. I. General theory of static properties," *J. Appl. Phys.*, vol. 44, pp. 1837-1849, doi :10.1063/1.1662460.

**Figure captions**

FIG. 1. (a) *M−H* curves for sample A at 300 K under in-plane and out-of-plane magnetic fields. (b) Temperature dependence of saturation magnetization, $M_s$, for samples A and B (circles and triangles, respectively). (c) Temperature dependence of perpendicular magnetic anisotropy energy density, *K*, for samples A and B (circles and triangles, respectively).

FIG. 2. MOKE microscope images of domain structures in sample A ((a)-(c)) and sample B ((d)-(f)) in demagnetized state at different temperatures (10, 200, and 300 K).

FIG. 3. Two-dimensional Fourier transforms of domain images for sample A at (a) 10 K and (b) 300 K.

FIG. 4. (a) Amplitude of direction-independent Fourier components obtained by averaging of the FFT of Fig. 3(b) over circles of radius 1/wavelength. (b) Schematic drawing of cross section of the assumed perpendicular regular stripe domain pattern. Temperature dependence of domain period $D_p$ for (c) sample A and (d) sample B obtained from domain images in demagnetized state, respectively. (e) Temperature dependence of domain wall surface energy $\gamma_w$ determined by the fit of the Kaplan and Gehring equations to (c) and (d) (circles and triangles correspond to sample A and B, respectively).

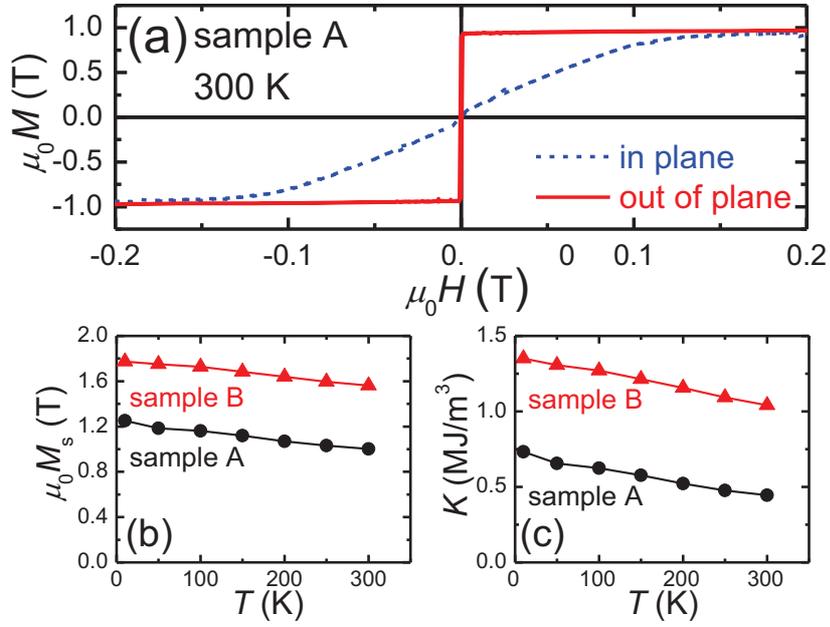

Figure 1

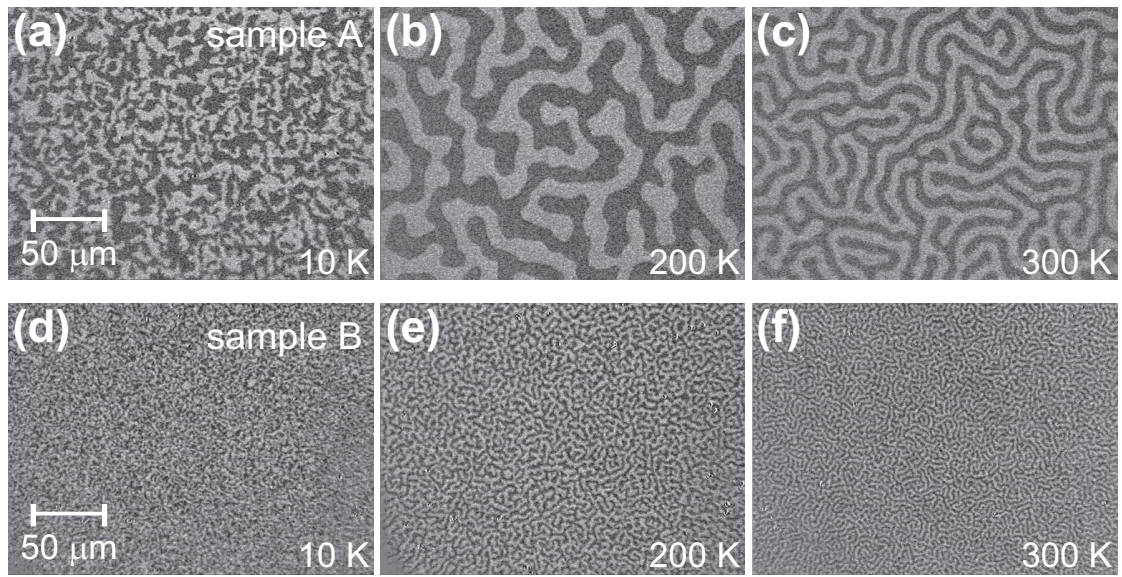

Figure 2

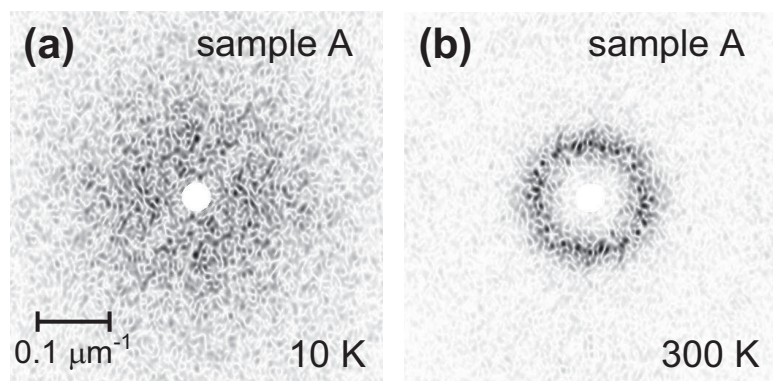

Figure 3

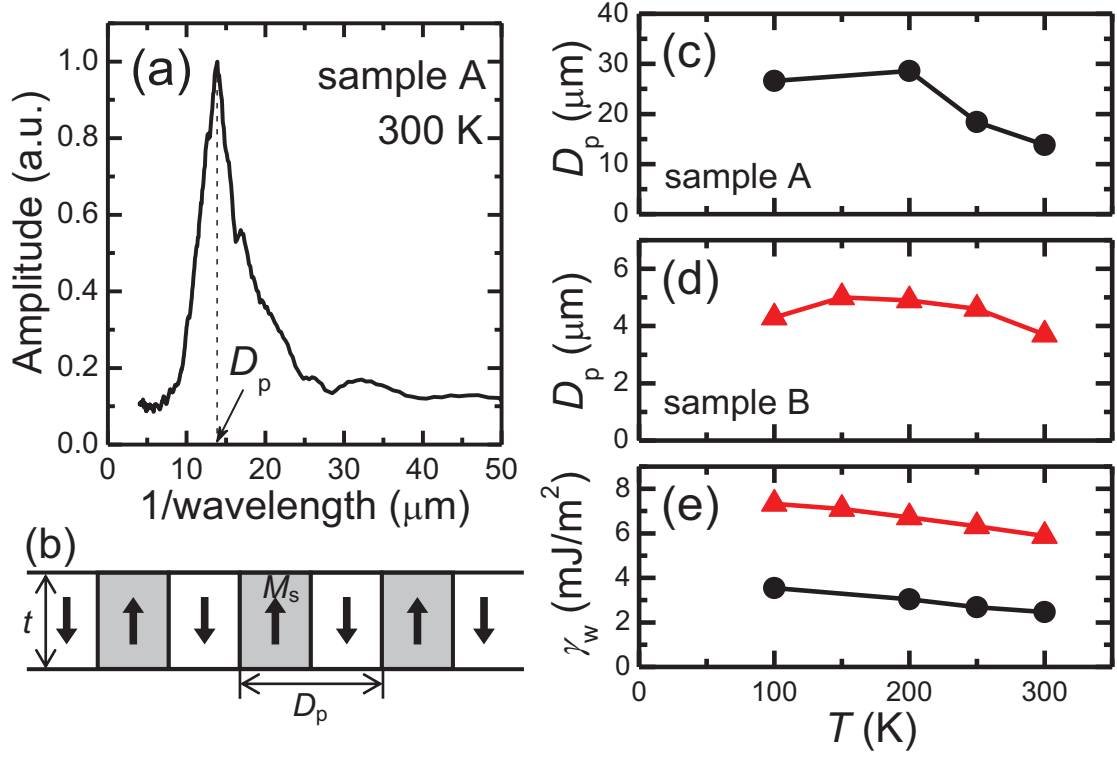

Figure 4